\documentclass[12pt]{article}

\usepackage{amssymb,epsfig}

\begin{document}

\begin{center}

{\Large \bf The NAIAD experiment for WIMP searches at Boulby mine and 
recent results}
\vspace{0.5cm}

{\large B. Ahmed~$^a$, G. J. Alner~$^b$, H. Araujo~$^a$, 
J. C. Barton~$^c$, A. Bewick~$^a$, \\ 
M. J. Carson~$^d$, D. Davidge~$^a$, J. V. Dawson~$^a$, T. Gamble~$^d$, \\
S. P. Hart~$^b$,
R. Hollingworth~$^d$, A. S. Howard~$^a$, W. G. Jones~$^a$, \\
M. K. Joshi~$^a$, 
V. A. Kudryavtsev~$^d$~\footnote{Corresponding author, 
e-mail: v.kudryavtsev@sheffield.ac.uk}, T. B. Lawson~$^d$,
V. Lebedenko~$^a$, \\
M. J. Lehner~$^d$~\footnote{Now at the University of Pennsylvania, Philadelphia, 
PA 19104, USA}, J. D. Lewin~$^b$, P. K. Lightfoot~$^d$, 
I. Liubarsky~$^a$, \\
R. L\"uscher~$^a$, J. E. McMillan~$^d$, B. Morgan~$^d$, 
G. Nicklin~$^d$, \\
S. M. Paling~$^d$, R. M. Preece~$^b$, J. J. Quenby~$^a$,
J. W. Roberts~$^b$, \\
M. Robinson~$^d$, N. J. T. Smith~$^b$, P. F. Smith~$^b$,
N. J. C. Spooner~$^d$~\footnote{Corresponding author, 
e-mail: n.spooner@sheffield.ac.uk}, \\
T. J. Sumner~$^a$, D. R. Tovey~$^d$}

\vspace{0.5cm}
$^a$ {\it Blackett Laboratory, Imperial College of Science, 
Technology and Medicine, London SW7 2BZ, UK}

$^b$ {\it Particle Physics Department, 
Rutherford Appleton Laboratory, Chilton, Oxon \\
OX11 0QX, UK}

$^c$ {\it Department of Physics, Queen Mary, University of 
London, London E1 4NS, UK}

$^d${\it Department of Physics and Astronomy, 
University of Sheffield, Sheffield S3 7RH, UK}

\vspace{0.5cm}
\begin{abstract}
The NAIAD experiment (NaI Advanced Detector) for WIMP
dark matter searches at Boulby mine (UK) is described. 
The detector consists of an array of encapsulated and 
unencapsulated NaI(Tl) crystals with high light yield.
Six crystals are collecting data at present. 
Data accumulated by four of them (10.6 kg$\times$year
exposure) have been used to set upper limits on the 
WIMP-nucleon spin-independent and WIMP-proton spin-dependent
cross-sections. Pulse shape analysis has been applied
to discriminate between nuclear recoils, as may be caused
by WIMP interactions, and electron recoils due to
gamma background. Various calibrations of crystals are
presented. 
\end{abstract}

\end{center}

\vspace{0.5cm}
\noindent {\it Key words:} Scintillation detectors, Inorganic crystals,
Dark matter, WIMP, Pulse shape analysis

\noindent {\it PACS:}  29.40.Mc, 14.80.Ly, 23.60.+e, 95.35.+d, 95.30.Cq

\vspace{0.5cm}
\noindent Corresponding authors: V. A. Kudryavtsev and N. J. C. Spooner, 
Department of Physics and Astronomy, University of Sheffield, 
Hicks Building, Hounsfield Rd., 
Sheffield S3 7RH, UK

\noindent Tel: +44 (0)114 2224531; \hspace{2cm} Fax: +44 (0)114 2728079; 

\noindent E-mail: v.kudryavtsev@sheffield.ac.uk, n.spooner@sheffield.ac.uk

\pagebreak

{\large \bf 1. Introduction}
\vspace{0.3cm}

\indent The UK Dark Matter Collaboration (UKDMC) has been operating 
NaI(Tl) detectors at the Boulby Mine 
underground site for several years \cite{ukdmc2}. 
Limits on the flux of weakly interacting massive 
particles (WIMPs), that may 
constitute up to $90\%$ of the mass of the Galaxy, have been published
using the data from
the first encapsulated detector \cite{ukdmc1,ukdmc3}. 
Pulse shape analysis (PSA) has been applied to the data
to distinguish between slow scintillations arising 
from background electron recoils and fast scintillations 
due to nuclear recoils, which are expected from WIMP-nucleus 
interactions \cite{dan1}. Improved limits were then obtained by 
the DAMA experiment, which also used NaI(Tl) scintillation 
detectors with PSA and had larger statistics \cite{dama1}.

Since then, the DAMA group moved to a simple annual modulation 
analysis of the background rate in its crystals without using PSA. 
The group claims that it observes an annual modulation consistent
with the expected signal from WIMP-nucleus interactions with
a specific set of WIMP parameters \cite{dama}.

Although several existing experiments have a potential to
probe the whole region of WIMP parameters allowed by the
DAMA signal (see, for example, 
Refs. \cite{cdms,edelweiss,cresst,zeplin},
and Ref. \cite{neil} for recent review), 
they use other techniques and other target
materials. This leaves room for speculation about
possible uncertainties in the comparison of results.
These uncertainties are related to systematic effects and 
nuclear physics calculations. 
Running an experiment, NAIAD, 
with the same target (NaI) and detection technique but different analysis
would help in the understanding of possible systematic effects. 
Such an experiment
will also be complemetary to more sensitive detectors in studying
regions of WIMP parameter space favoured by the DAMA positive
signal.

The advance of the UKDMC NaI(Tl) experiment has been blocked
for a few years by the discovery of a fast anomalous
component in the data from several encapsulated crystals
\cite{ukdmc2}. These events were faster than typical electron 
recoil pulses and faster even than nuclear recoil pulses 
\cite{ukdmc2,vak}. 
Similar events have also been seen by
the Saclay group \cite{gerbier}.
The most plausible explanation
of these events at the moment is implanted surface 
contamination of the crystal by an alpha-emitting
isotope from radon decay \cite{smith,csitest}. 

NAIAD was constructed with unencapsulated crystals
to allow better control of the crystal surface, to reduce
the background due to such events and to improve
the light collection \cite{naiad}. 
Such an experiment could probe
the region of WIMP parameters allowed by the DAMA experiment
using a similar detection technique and target,
and reach the region favoured by some SUSY models \cite{naiad}.
NaI has an advantage of having two targets with high and low
masses, thus reducing uncertainties related to nuclear
physics calculations. The detectors are sensitive to both
spin-independent and spin-dependent interactions. The
experiment is complementary to other dark matter experiments
at Boulby, such as ZEPLIN \cite{zeplin} (xenon gas and double-phase
xenon detectors) and DRIFT \cite{drift} (time projection chamber
with directional sensitivity).
The array of NaI(Tl) detectors can also be used as a
diagnostic array to study background and systematic effects
for other experiments at Boulby.
(Note that the new data from this array are indeed free
from an anomalous fast component \cite{csitest}).

In this paper we present the status of the NAIAD array 
(NaI Advanced Detector) at Boulby. Section 2
describes the experimental set-up. The analysis procedure, 
various calibrations and their results are presented in Section 3. 
WIMP limits from data collected up to the end of 2001 are shown
in Section 4. Our conclusions are given in Section 5.

\vspace{0.5cm}
{\large \bf 2. NAIAD experiment}
\vspace{0.3cm}

The NAIAD array is operational in the underground laboratory
at Boulby mine (North Yorkshire, UK) at a vertical depth
of about 1100 metres.
In its final stage the NAIAD array will consist of eight
NaI(Tl) crystals from different manufacturers (Bicron, Hilger,
VIMS). At present (May 2002) 6 detectors are running with a
total mass of 46 kg. Two detectors contain encapsulated
crystals, while 4 other crystals are unencapsulated.
To avoid their degradation by humidity in the atmosphere, the
unencapsulated crystals have been sealed in copper boxes filled
with dry nitrogen. 
The UKDMC moved away from the original proposal to run the crystals
in pure mineral oil \cite{naiad} because of an observed slow 
degradation of one of the crystals with that system.

\begin{figure}[htb]
\begin{center}
\epsfig{figure=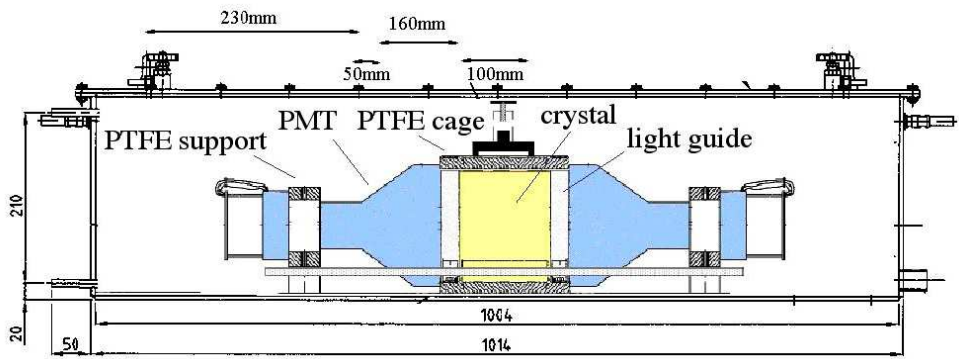,height=5cm}
\caption {Schematic view of one NAIAD module.}
\label{det}
\end{center}
\end{figure}

A schematic view of one detector is shown in Figure \ref{det}.
A crystal (in the middle) is mounted in a 10 mm thick solid
PTFE (polytetrafluoroethylene) reflector cage and is coupled 
to light guides. The
two 4-5 cm long quartz light guides are also mounted in the 
PTFE cages and are coupled to 5 inch diameter
low background photomultiplier tubes (PMTs), ETL type 9390UKB.
Only selected low background materials are used in the detector design
including oxygen-free high-conductivity copper and PTFE.

Temperature control of the system is achieved through copper coils
outside the copper box. Chilled water is constantly pumped 
through the coils maintaining the temperature of the crystals 
at $(<T> \pm 0.1)^{\circ}$C during a single run, where $<T>$
depends on the crystal and is typically about $10^{\circ}$C.
The temperature of the crystal,
ambient air, water in the pipes and copper is measured
by thermocouples. Although the temperature differs
from one crystal to another, for any particular crystal it remains
stable within $\pm 0.1^{\circ}$C during a single run. If, for any
reason (for example, chiller failure), the variation of crystal 
temperature exceeds the predefined limit, the data from these periods
are not included in the analysis. If any changes in the experimental
set-up result in a change of the mean temperature of the crystal,
the data from different runs are not combined together, but assumed
to come from different experiments, so only the resulting limits
can be combined.

Pulses from both PMTs 
are integrated using a buffer circuit and then digitised using a 
LeCroy 9350A oscilloscope driven by 
a Macintosh computer running Labview-based data acquisition (DAQ) 
software. 
The digitised pulse shapes (5 $\mu$s total digitisation time, 
10 ns digitisation accuracy) 
are passed to the computer and stored on disk. 
The gain of the PMTs is set to give about 2.5 mV per 
photoelectron. This has been found to be the optimum between the high gain
required by digitisation and noise level, and low gain to avoid
afterpulses.
Low threshold discriminators are typically set to 10 mV threshold,
which corresponds to about 4 photoelectrons (pe) or about 1 keV for
the crystal with total light yield of 8 pe/keV.

Copper boxes containing the crystals are installed in 
lead and copper ``castles'', to shield the detectors from
background due to natural radioactivity in the surrounding rock. 
The data reported here were taken without neutron shielding.
Wax and polypropylene neutron shielding around the castles 
has now been installed, thus improving
the sensitivity of the experiment to WIMP interactions.

The temperature of the crystals is monitored by the ``slow control'' part
of the data acquisition. Crystal temperature and ambient temperature
are recorded in the data stream for each event. The temperature of
the lead in the castle and water in the pipes is stored in
``log'' files. The slow control part of the DAQ is also responsible for
daily calibration of the detectors with gamma sources, described
in Section 3.

The light yield of the crystals is obtained from measurement of
the single photoelectron pulses (about 2.5 mV after electronic noise
subtraction) and the 122 keV $\gamma$-peak from $^{57}$Co.
The six crystals currently collecting data have light yield from 4.6 up to
9 pe/keV. The light yield is
checked every 2-4 weeks and is found to be stable within 10\% for
all crystals. The
longest operated crystal, running since the summer of 2000, shows
a degradation of no more than 10\% over the whole operation period.

After the first period reported here,
the data acquisition was changed from the oscilloscope based system
described above to one based on fast Acqiris CompactPCI
digitisers. This has allowed significant reduction in the dead time
of the data acquisition, which is particularly important for
high rate calibration runs. The cost of DAQ hardware has been
reduced dramatically, since several detectors can now be controlled 
by one computer, and high-cost oscilloscopes have been substituted
by Acqiris digitisers.

\vspace{0.5cm}
{\large \bf 3. Analysis procedure and calibrations}
\vspace{0.3cm}

Final analysis has been 
performed on the sum of the pulses from the two PMTs attached to each crystal.
The parameters of the pulses from each PMT have been used to
apply so-called asymmetry cuts (described below) to remove those events with
obvious asymmetry between the pulses from each PMT.
Our standard procedure of data analysis
involves fitting of a single 
exponential to each integrated pulse in order 
to obtain the index of the exponent, 
$\tau$, the amplitude of the pulse, $A$, and the start time, $t_s$.
The scintillation pulses from nuclear and electron recoils have been
shown to be nearly exponential in shape \cite{dan1}.
In addition to these, the mean time of the pulse (mean photoelectron 
arrival time), as an estimate
of the time constant, $\chi^2$ of the fit, number of photoelectrons and 
the energy are also calculated for each pulse. The conversion of the 
pulse amplitude to the number of photoelectrons and the energy is done
using pre-determined conversion factors, which come from
energy and single photoelectron calibrations. These calibrations
are performed typically every 2-4 weeks to ensure that there
are no changes in the PMT gains or crystal degradation. (Note that
the monitoring of the temporal behaviour of the energy spectrum 
in most cases can also reveal if changes have occurred in the system).
Energy calibration is performed with $^{57}$Co gamma-ray source
(122 keV line). 

For each run (or set of runs) the ``energy -- time constant''
($E-\tau$) distribution is constructed. If all operational settings 
(including temperature) are the same for several runs, the
($E-\tau$) distributions for these runs are summed together.
Time periods with temperature of the crystal outside a predefined range
have been removed from the analysis. Also all events with
$\tau$ less than 20 ns in either PMT are excluded.

To reduce PMT noise and, particularly, events where
a spark (flash) in the dynode structure of one PMT \cite{krall}
is seen by both PMTs, the so-called ``asymmetry cuts'' are
applied. These cuts are based on Poisson statistics and use
the asymmetry in amplitude,
time constant and start time of noise (non-scintillation) pulses 
from the two PMTs to remove them.

For any small energy bin (1 keV width, for example), the time
constant distribution can be approximated by a Gaussian in 
$\ln(\tau)$ ($\log$(Gauss) function) \cite{ukdmc1,vak} (for a more detailed 
discussion of the distributions see \cite{dan2} and references therein):

\begin{equation}
{{dN}\over{d\tau}} = {{N_o}\over{\tau \sqrt{2\pi} \ln w}} \times
\exp \Big[ {{-(\ln \tau-\ln \tau_o)^2}
\over{2(\ln w)^2}} \Big]
\label{logg}
\end{equation}

The $\tau$-distributions are fitted with a 
Gaussian in $\ln(\tau)$ 
with three free parameters: mean time constant
$\tau_o$, width $w$ and normalisation factor $N_o$. 
In experiments where a second population
is seen (for example, nuclear recoils from a neutron 
source or possible WIMP-nucleus interactions), the resulting 
$\tau$-distribution can be fitted with 
two $\log$(Gauss) functions with the same width $w$
(we assume the same width for both populations since
the width is determined mainly by the number of collected 
photoelectrons). 

The aim of the analysis procedure is to find the
second population of events in $\tau$-distributions or
to set an upper limit on its rate. To reach this,
the response of the crystals to various radiations
should be known. This is achieved through the
measurements of the $\tau$-distributions for all energies of 
interest (2-40 keV) with gamma-ray ($^{60}$Co) and neutron
($^{252}$Cf) sources. Photons from high-energy
gamma-ray source produce Compton electrons in the crystal
volume similar to those initiated by gamma-ray background at
Boulby. Neutrons collide elastically with the nuclei of the
crystal giving nuclear recoils similar to those expected
from WIMP-nucleus elastic scattering.
Calibration experiments were
performed on all crystals at the surface prior to moving the
crystals underground. Although the temperature was
not maintained exactly the same as the operational temperature
in the mine, this was found to have no effect on the
ratio of the mean time constants of nuclear and
electron recoils, $R_{\tau}=\tau_n/\tau_e$. 

\begin{figure}[htb]
\begin{center}
\epsfig{figure=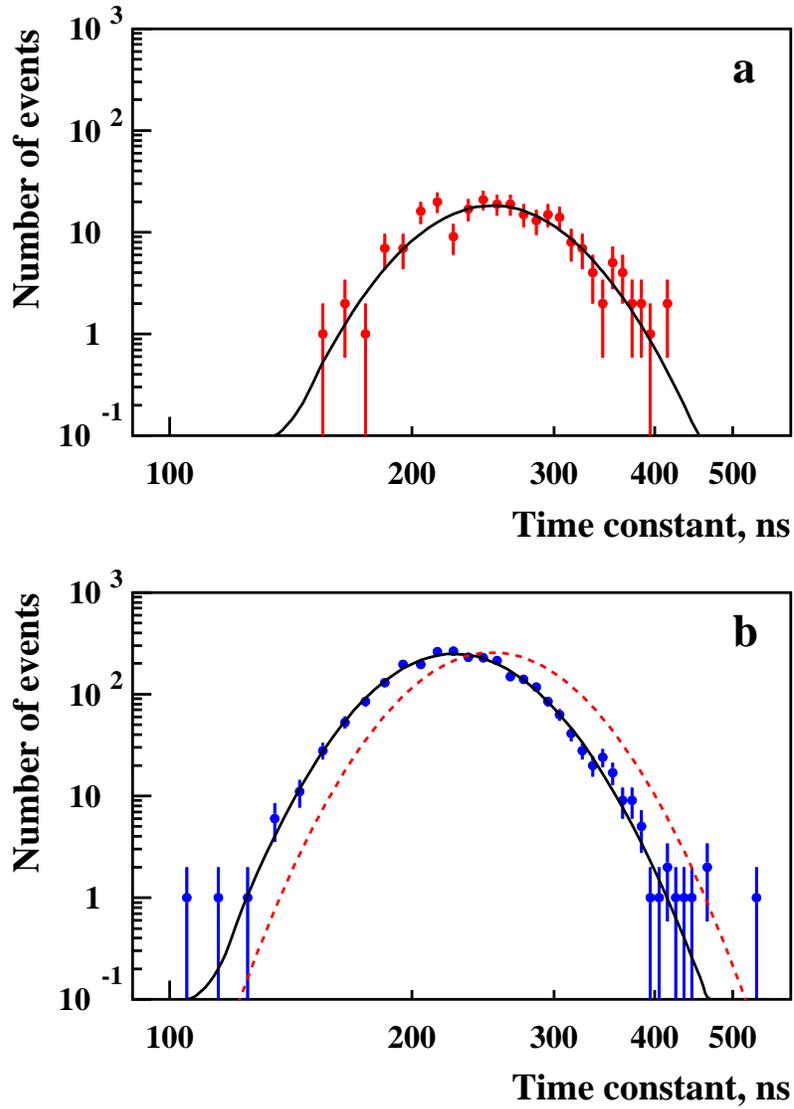,height=15cm}
\caption {a) Time constant distribution for Compton electrons 
with visible energy
6-8 keV from high-energy gamma-ray source;
b) similar distribution for nuclear recoils from neutron source.
Solid curves show fits to Gaussian distributions
on a logarithmic scale ($\log$(Gauss)-function). 
Dashed curve on (b) is the fit for electron recoils from (a)
normalised to the peak value for nuclear recoils.}
\label{tau}
\end{center}
\end{figure}

\begin{figure}[htb]
\begin{center}
\epsfig{figure=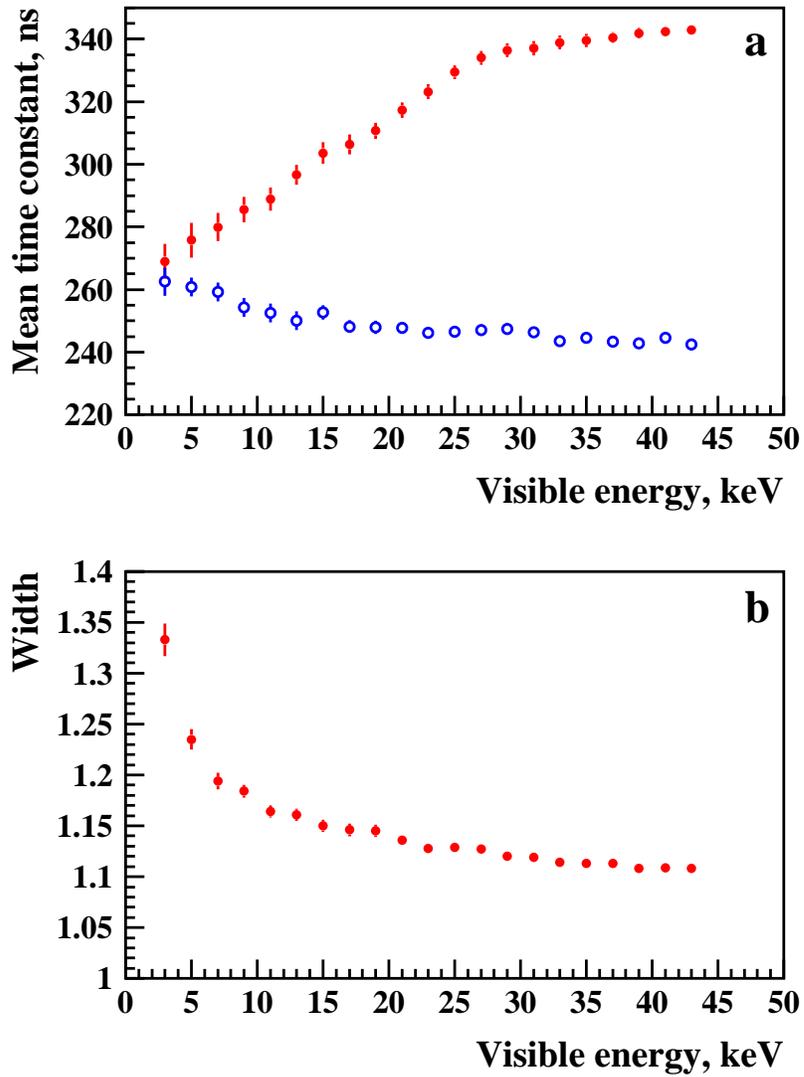,height=15cm}
\caption {a) Mean time constant for electron (filled circles)
and nuclear (open circles) recoils as a function of visible energy in DM77;
b) width of the $\log$(Gauss)-fit as a function of visible energy.}
\label{tau-w}
\end{center}
\end{figure}

Figure \ref{tau}
shows the $\tau$-distributions of events in one of the crystals
(DM77) collected with
gamma-ray (a) and neutron (b) sources for 7-8 keV visible
energy. Figure \ref{tau-w} shows the mean time constants
of electron and nuclear recoils (a)
and the width of the $\log$(Gauss) fit (b) as functions of energy.
As can be seen from Figure \ref{tau-w} the discrimination 
between nuclear and electron recoils is possible only
at visible energies more than 4 keV. However, for
crystals with high light yield, the width, $w$, is small 
and the discrimination power is significant even at low
energies \cite{naiad}. Our measurements of the time constant
as a function of energy are in agreement with the results
reported in Ref. \cite{gerbier}.

Another critical feature of the detector response is the energy 
resolution. It is important in the procedure of setting limits 
on the rate of nuclear recoils produced by WIMP-nucleus 
interactions. The procedure requires calculation of the recoil 
spectra as functions of visible energy for various WIMP masses 
and their comparison with the measurements \cite{ls}. A recoil 
spectrum calculated in a particular model should be folded 
with the detector response function to obtain a visible
energy spectrum. The detector response function in this case
is the energy resolution function, which gives the probability
distribution of the deposited energy being seen by a detector
as a certain visible energy. Assuming the probability
distribution is a Gaussian function (for a Poisson process
with large number of photoelectrons), the energy resolution
is characterised by the width, $\sigma$, of the Gaussian
function or by the full width at half maximum of the 
distribution, FWHM (the latter is applicable to any form of
the detector response function). 

The energy resolution of the NAIAD detectors 
is normally measured during the standard
procedure of energy calibration with a $^{57}$Co
source (122 keV line) performed every 2-4 weeks. However, 
nuclear recoils from WIMP-nucleus interactions can be
seen mainly at low visible energies (4-30 keV). Ideally, 
we need to perform the measurements of energy resolution
as a function of energy at low energies. However, practically
this is impossible because we cannot access the surface
of the crystals sealed in the copper boxes
during the experiment, and
low energy photons will be absorbed in the copper or other 
materials surrounding the crystal. Moreover, even for 
unencapsulated crystals, any calibration
with low energy photons refers only to the surface area close
to the gamma-ray source (due to the high photon absorption).
The uniformity of the crystal is essential for the use of
such a calibration.

\begin{figure}[htb]
\begin{center}
\epsfig{figure=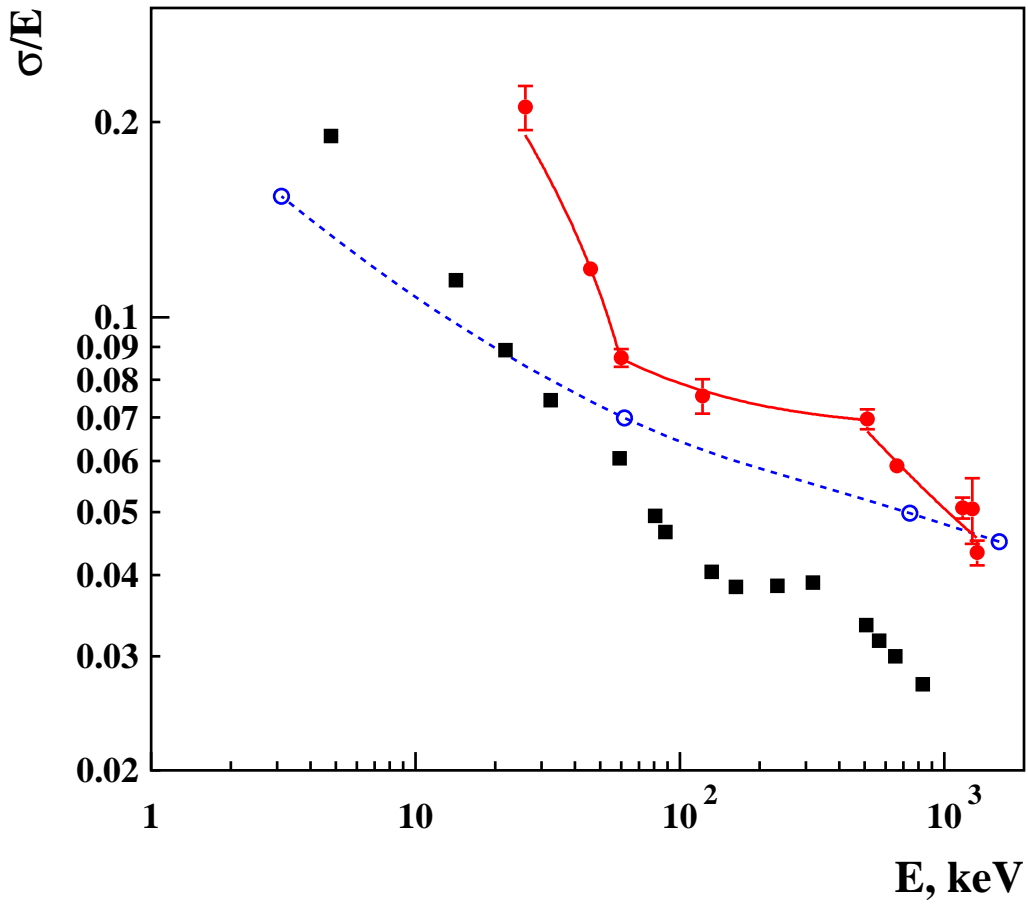,height=13cm}
\caption {Energy resolution of the UKDMC detector DM77 (filled circles
with the fit),
DAMA detector (open circles with smooth curve drawn through the 
experimental points) 
\cite{damares} and small size crystal (filled squares)
\cite{sakai}.}
\label{res}
\end{center}
\end{figure}

Prior to moving the detectors underground the energy resolutions
of the crystals at various energies were measured with a number
of gamma-ray sources. Figure \ref{res} shows the width of the 
Gaussian fit to the measured gamma-ray line as a function of
photon energy (filled circles) for one of the UKDMC crystals (DM77). 
The gamma-ray sources were
attached to the crystal surface during the measurements. For
low-energy sources, when the photons are absorbed within a few
millimetres or less, the dependence of the energy
resolution on the source position was investigated. 
The energy resolution of the sum of the pulses
from the two PMTs was found to be independent of the source position, thus
confirming the uniformity of the crystal. (Note, however, that
this conclusion refers to the crystal surface only, since 
low-energy photons cannot penetrate deeply into the crystal volume).

The data presented in Figure \ref{res} cannot be fitted with
a single function. Our measurements agree reasonably well with
the measurements by Sakai \cite{sakai}, in which a small size
($1 \times 1 \times 2$ cm$^3$) NaI(Tl) crystal was used. 
The data from Ref. \cite{sakai} are shown by filled squares in Figure
\ref{res} (FWHM has been converted
to the width of Gaussian fit using the relation:
$\sigma$=FWHM/2.35). 
Both measurements (UKDMC and Ref. \cite{sakai}) reveal a
complicated dependence of the energy resolution on the energy.
Three regions with different slopes are clearly seen. Similar
effects were reported earlier (see, for example, Ref. \cite{birks}
and references therein), although for a restricted energy range.
Note that the resolution of our crystal is worse than the one
reported in Ref. \cite{sakai} due to the much larger crystal used.

According to the theory of scintillation counting (for discussions
see Refs. \cite{birks,dorenbos}) the energy dependence of the
resolution function is approximated as:

\begin{equation}
\Big ( {{\sigma} \over {E}} \Big) ^2 = a + {{b} \over {E}}
\label{enres}
\end{equation}

The parameters $a$ and $b$ have been determined from the best fits
to the data in three energy regions, as shown in Figure \ref{res}.
This procedure has been repeated for all crystals. The energy
resolution has been extrapolated to lower energies (4-20 keV) using 
Eq. (\ref{enres}) with the parameters for low energy region (26-60 keV).

Also shown in Figure \ref{res} are the reported measurements of
energy resolution by the DAMA group \cite{damares} (open circles,
dashed curve is drawn through the centres of the points).
At high energies the resolution of DAMA crystals is similar to ours, 
while at low energies it is much better than ours and
even better than that of the much smaller crystals
(filled circles from Ref. \cite{sakai}). The energy dependence
of the resolution of the DAMA crystals is different from our
measurements and those of Ref. \cite{sakai}. The lowest energy
at which the resolution was measured by the DAMA group, is a little 
more than 3 keV, but the authors of Ref. \cite{damares} 
did not specify the source of this line. 
At 4 keV (the minimal
energy at which we observed discrimination between electron and 
nuclear recoil) a typical resolution of NAIAD detectors
is about 0.5 (0.56 for DM77 shown in Figure \ref{res}), a factor of
3-4 poorer than in the DAMA experiment.
The resolution given by DAMA for their crystals is even better than
the theoretical limit, determined by the light yield, which
contradicts the basic theory of scintillation counting \cite{birks}:
for a light yield of about 6 pe/keV \cite{dama-lightyield},
the resolution limit at 3.2 keV
(the first point on the graph \cite{damares}) is 
$1/\sqrt (3.2 \times 6) \approx 0.23$ -- poorer than the
value reported by DAMA ($\approx 0.15$ \cite{damares}).

\vspace{0.5cm}
{\large \bf 4. Results and discussion}
\vspace{0.3cm}

The UK Dark Matter Collaboration has been operating the
detectors of the NAIAD array since the winter of 2000. The first
crystal (DM74) was immersed in pure mineral oil and ran 
for more than six months until degradation of the crystal 
surface resulted in a 
significant decrease of the light yield. The second crystal
(DM72) was installed in summer of 2000 in a sealed
copper box and has been running since then with a loss of light 
yield of no more than 10\%. The data collected with these
two crystals have been used to set preliminary limits
on the WIMP-proton spin-dependent and WIMP-nucleon
spin-independent cross-sections \cite{york,icrc}.

More crystals were added to the array in 2001. 
At present six detectors
are running with a total mass of 46 kg. 
None of the 6 crystals show anomalous fast events.
Two more crystals are expected to be added in 2002. The final mass
of the array will be about 65 kg. 

The data from 4 crystals have been used to set the limits
on WIMP-nucleus cross-section reported here.
The total statistics include five runs, with one of the
crystals (DM74) running twice: in mineral oil and
in dry nitrogen (after re-polishing). 
Table \ref{stat} shows main characteristics and statistics for
all detectors.

The software energy threshold has been set to 2 keV, while
the typical energy threshold of discriminators was about 1 keV.
An energy range from 2 to 30 keV has been used in the data
analysis. Pulse shape discrimination has been applied for
pulses exceeding 4 keV. Since there is no visible discrimination
between electron and nuclear recoils below 4 keV, the total
background rate has been used to set upper limits on the nuclear
recoil rate at 2-4 keV.

\begin{figure}[htb]
\begin{center}
\epsfig{figure=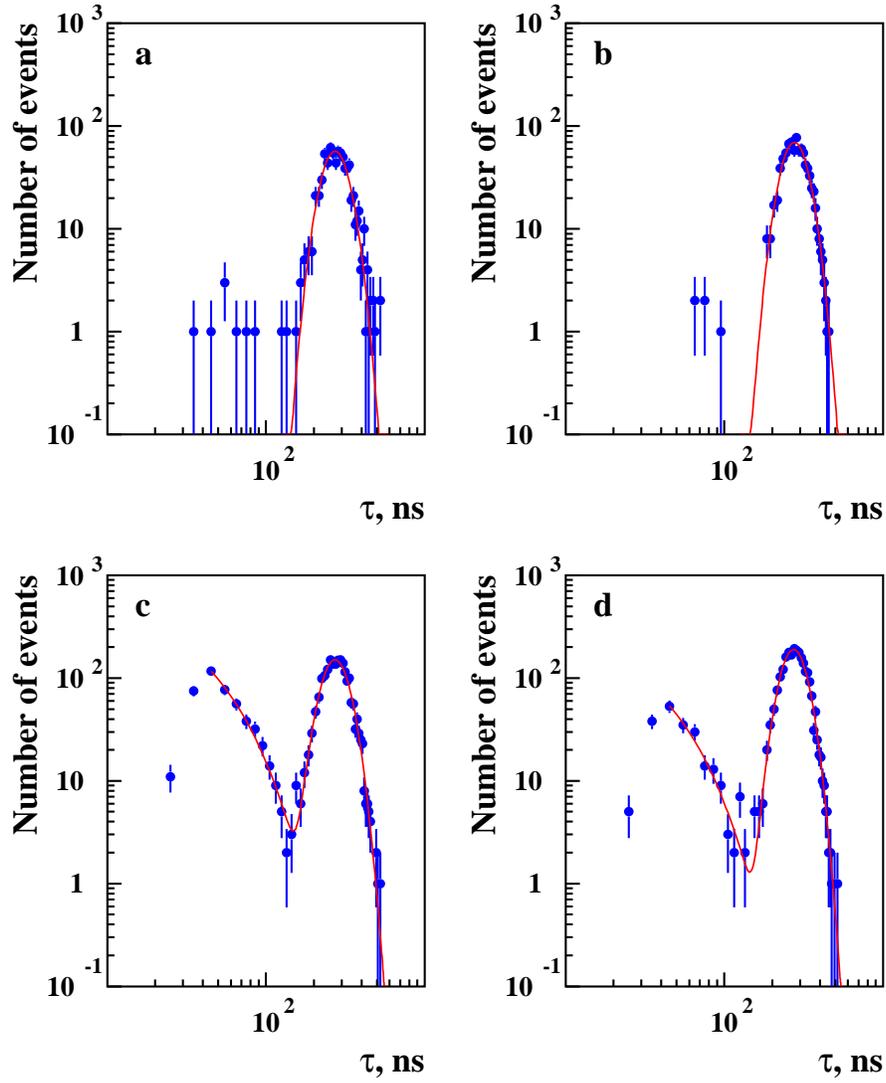,height=15cm}
\caption {Typical time constant distributions at energies 6-7 keV
(a,c) and 7-8 keV (b,d) together with the best fits:
(a) and (b) - calibration run with gamma-ray source $^{60}$Co (compton
calibration); (c) and (d) - data run. For calibration run
only a $\log$(Gauss) fit is applied. For data run an
exponential is added to fit the PMT noise.}
\label{tcd}
\end{center}
\end{figure}

Figure \ref{tcd} shows typical time constant
distributions at 6-8 keV from data and calibration runs.
The PMT noise events are seen at small values of time constants.
These events are fitted with an exponential, in addition to the
log(Gauss) fit to the scintillation pulses.
Figure \ref{tcd} does not reveal any visible difference
between data and calibration runs in terms of time constant
distributions (apart from the presence of PMT noise at small $\tau$).
The limit on the nuclear recoil rate at any particular energy
has been obtained by fitting the measured time constant distribution
with two $\log$(Gauss) functions having known parameters: mean
time constants and widths, known from Compton and neutron 
calibrations. An exponential fit to the PMT noise has been added
if necessary. Free parameters thus remaining were the total numbers
of electron and nuclear recoils. These numbers have been restricted
to non-negative values. The best fit numbers of nuclear recoils
have been found to be either zeros or small positive values.
The positive values have normally been within 1.5 standard deviations
from zero, implying that they were statistical fluctuations around
the mean value equal to 0. (Note that negative values have
no physical meaning and were not allowed in the analysis).
No statistically significant number of
nuclear recoils has been found for any energy bin or any crystal
under consideration. Thus, the upper limits on the nuclear
recoil rate have been obtained for each energy bin and for each
crystal. As was mentioned above, the visible energy spectrum
of nuclear recoils is the convolution of the calculated spectrum
with the detector response function. This means that there is no
model-independent procedure of unfolding the expected recoil 
spectrum from the visible spectrum. 

\begin{figure}[htb]
\begin{center}
\epsfig{figure=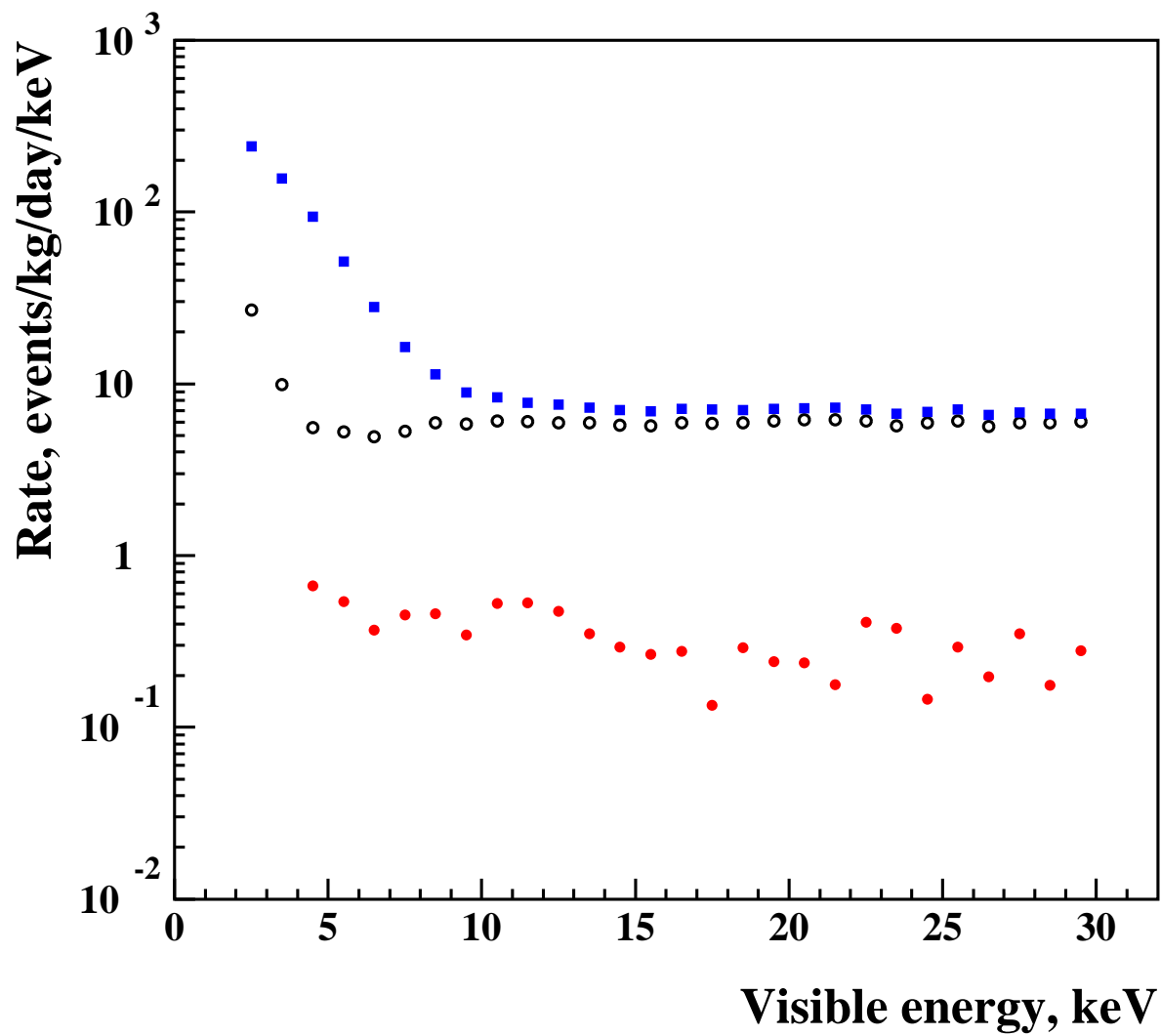,height=15cm}
\caption{Total energy spectrum from
one of the crystals (DM74) (filled squares), spectrum of electron recoils 
after asymmetry cuts from the fit to
the main peak on the time constant distribution (open circles), 
and the upper limits on the nuclear recoil rate for 
various energy bins (filled circles).}
\label{spectrum}
\end{center}
\end{figure}

Figure \ref{spectrum} shows a typical total energy spectrum from
one of the crystals, spectrum of electron recoils after asymmetry cuts
from the fit to
the main peak on the time constant distribution, and the 90\% C.L. upper
limits on the nuclear recoil rate for various energy bins.
The limit on the nuclear recoil rate
at 4-5 keV for our statistics is typically about (10-20)\% of the 
background rate due to the discrimination power of PSA. 
Note that the limit on the nuclear recoil rate presented in 
Figure \ref{spectrum}, when analysed in terms of WIMP-nucleon interactions,
is the convolution of the expected rate
from WIMP interactions and the detector response function. This
makes direct comparison of the limits on the rate from
different experiments difficult.
We found that for all our crystals 
the rate below 4 keV, where the discrimination cannot be applied,
does not contribute much to the limits on the WIMP-nucleus
cross-section. This is because of the increase in rate with
decrease in visible energy, that gives a factor of 20
difference in a residual rate (after discrimination) between 3-4
keV and 4-5 keV bins. This increase in residual rate is higher than the
difference in expected rate from WIMP interactions folded with our
typical detector response function. This makes the limit on the
cross-section weakly sensitive to the points below the disrimination
threshold.

\begin{figure}[htb]
\begin{center}
\epsfig{figure=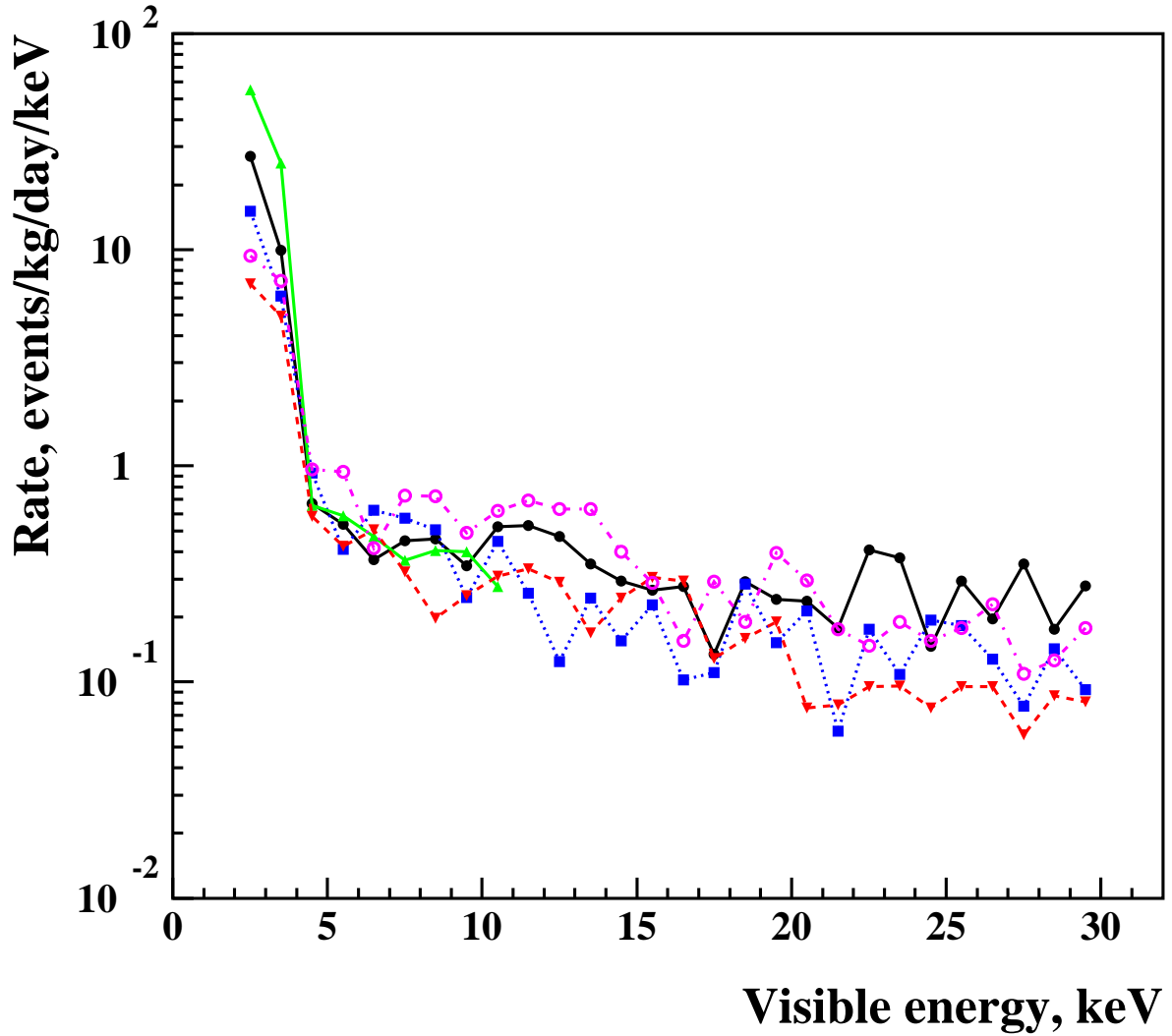,height=15cm}
\caption{Upper limits (90\% C.L.) on the nuclear recoil rate
from different crystals: DM74 in dry nitrogen --
filled circles and thick solid line connecting the points;
DM74 in oil -- filled squares and dotted line;
DM72 -- triangles pointing up and thin solid line 
(restricted energy range up to 11 keV was analysed);
DM76 -- triangles pointing down and dashed line;
DM77 -- open circles and dash-dotted line.}
\label{ratelimit}
\end{center}
\end{figure}

The 90\% C.L. limits on the nuclear recoil rate as a function of visible
(measured) energy are shown in Figure \ref{ratelimit} for
each crystal. 
The limits on the nuclear recoil rate for each energy bin and each crystal 
have been converted into limits on the WIMP-nucleon spin-independent
and WIMP-proton spin-dependent cross-sections following the 
procedure described by Lewin and Smith \cite{ls}. Expected
nuclear recoil spectra from WIMP-nucleus interactions have
been calculated for a halo model with parameters:
$\rho_{dm}$ = 0.3 GeV cm$^{-3}$, $v_o$ = 220 km/s, 
$v_{esc}$ = 650 km/s and $v_{Earth}$ = 232 km/s.
For the spin-independent case the form factors have been computed
using Fermi nuclear density distribution with the parameters
to fit muon scattering data as described in Ref. \cite{ls}.
The WIMP-nucleus scattering has been taken proportional
to $A^2$.
For the spin-dependent case a pure higgsino is assumed.
The spin factors and form factors have been computed
for sodium and iodine nuclei on the
basis of nuclear shell model calculations of Ressell and Dean
\cite{ressel} using their ``Bonn A potential'' results
for iodine.
The quenching factors (scintillation efficiencies)
have been taken as 0.275 for sodium and 0.086 for iodine
recoils \cite{dan1}.

The limits on the cross-section for various energy bins, targets
(sodium and iodine) and crystals have been combined
following the procedure described in Ref. \cite{ls}
assuming the measurements for different energy bins
and different crystals are statistically independent.
First, the limits on the cross-sections from various
energy bins (for each crystal) have been combined 
statistically using the equation: 

\begin{equation}
{{1}\over{\sigma_{ij}^2 (M_{W})}} = \sum_{k=1}^{k_{max}}
{{1}\over{\sigma_{ij}^2 (E_k,M_{W})}}
\label{enlimits}
\end{equation}

\noindent where $\sigma_{ij} (E_k,M_{W})$ is the limit on the
cross-section for WIMP-nucleon interaction (WIMP mass $M_{W}$)
at 90\% C.L. from the energy bin with central
energy $E_k$. The index $i$ denotes the crystal, while the index
$j$ denotes the nucleus (sodium or iodine).

Limits for sodium and iodine each ignore the contribution of the other
element; a better estimate of the combined limit from NaI is given by:

\begin{equation}
{{1}\over{\sigma_{i} (M_{W})}} = \sum_{j=1}^{2}
{{1}\over{\sigma_{ij} (M_{W})}}
\label{ellimits}
\end{equation}

\noindent Note that the fraction of each element by weight was taken
into account in the calculation of the interaction rate for each element.

Finally, the limits from different crystals have been combined
using the equation below, taking into account that the data were
statistically independent and no positive signal was detected
in any of the crystals:

\begin{equation}
{{1}\over{\sigma^2 (M_{W})}} = \sum_{i=1}^{i_{max}}
{{1}\over{\sigma_{i}^2 (M_{W})}}
\label{totlimits}
\end{equation}

\begin{figure}[htb]
\begin{center}
\epsfig{figure=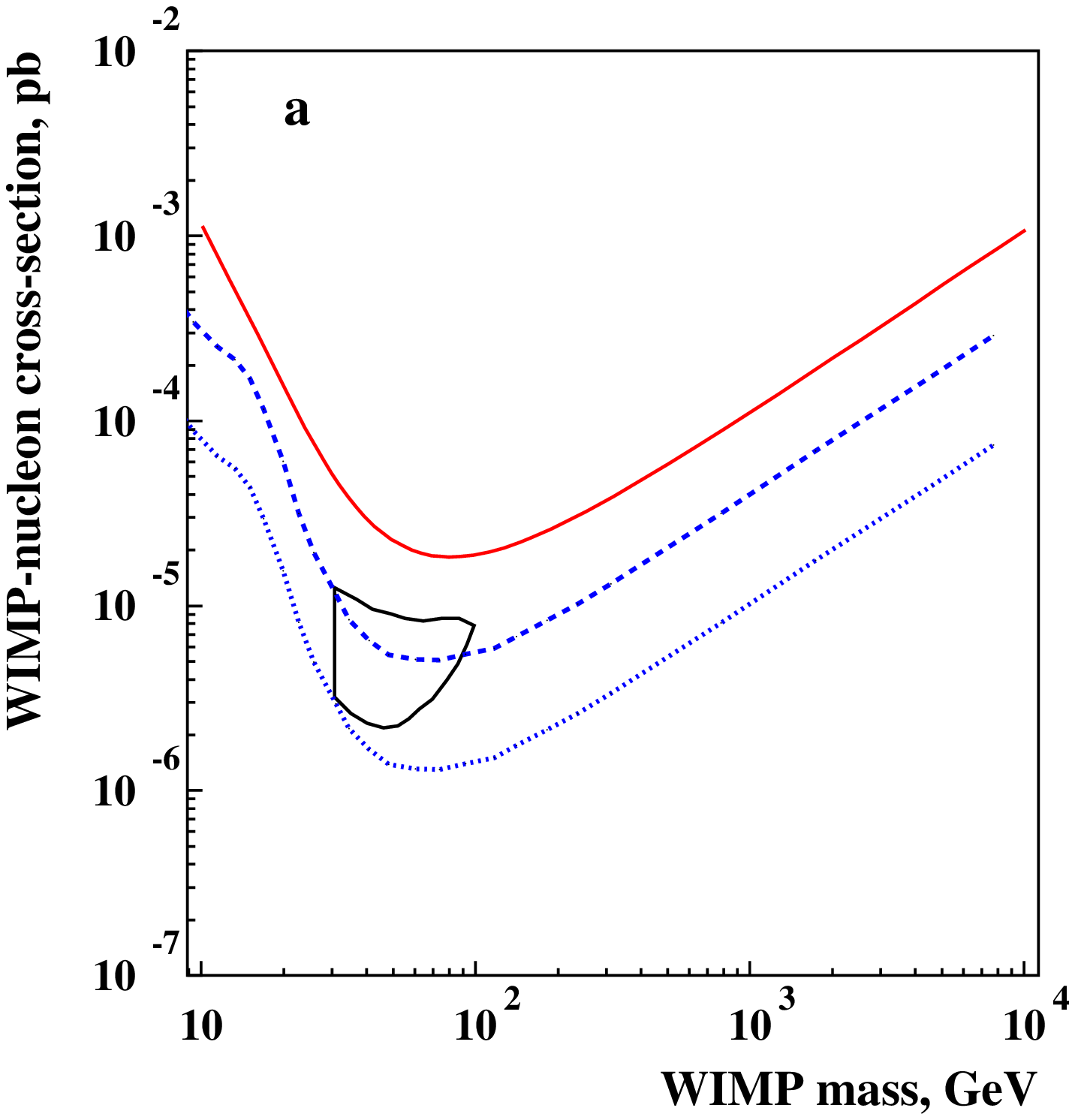,height=9cm}
\epsfig{figure=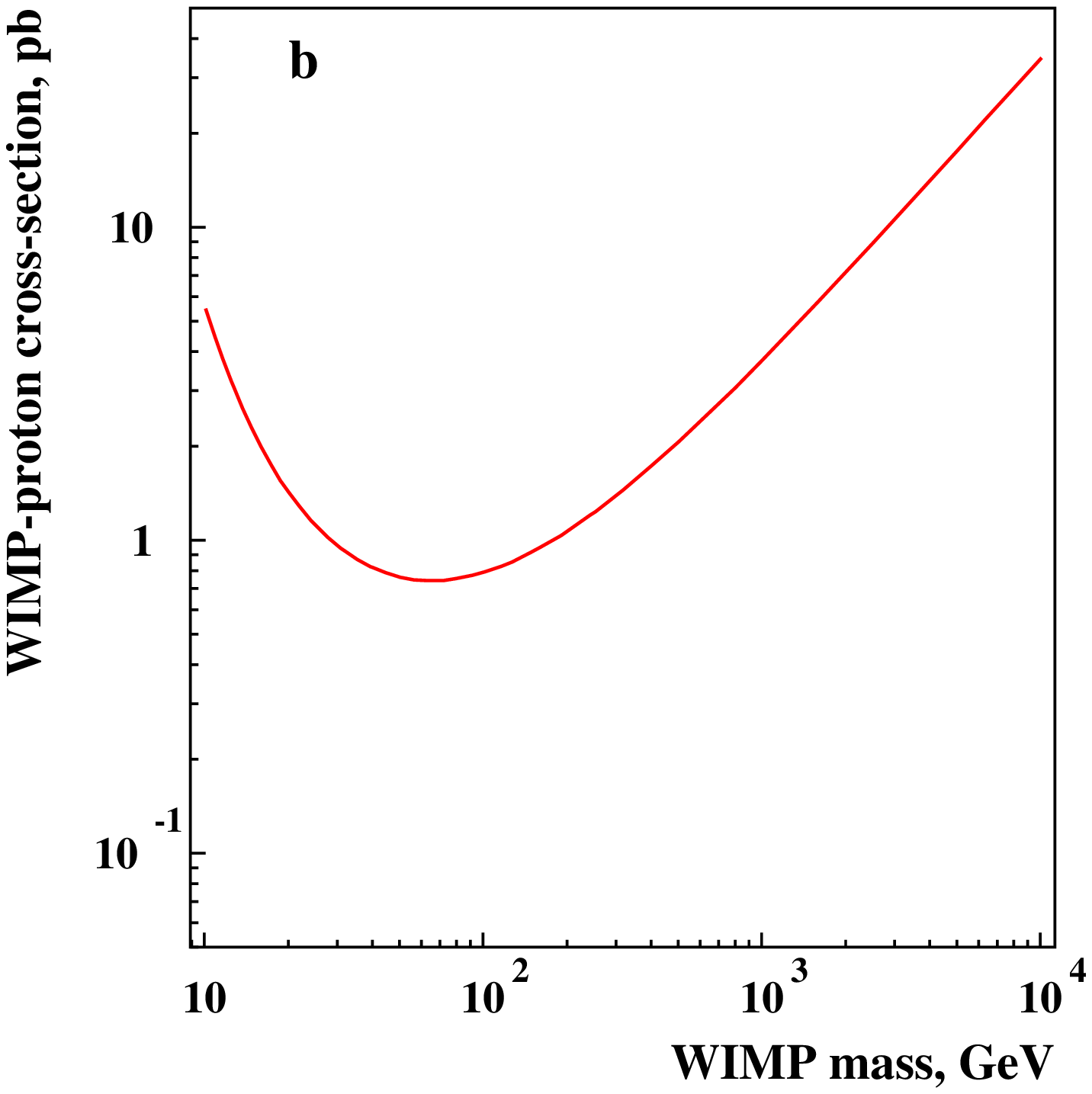,height=9cm}
\caption {NAIAD limits (90\% C.L.)
on WIMP-nucleon spin-independent (a) and WIMP-proton spin-dependent
(b) cross-sections as functions of WIMP mass.
Also shown are the 
region of parameter space favoured by the DAMA positive
annual modulation signal (DAMA/NaI-1 through DAMA/NaI-4) 
(closed curve), limits on the spin-independent cross-section set by the DAMA
experiment (DAMA/NaI-0) using pulse shape analysis (dashed curve) and
the projected limit of DAMA experiment (DAMA/NaI-0 through
DAMA/NaI-4, dotted curve), if the DAMA group were to apply pulse shape
discrimination to all available data sets.}
\label{limits}
\end{center}
\end{figure}

Figure \ref{limits}a (b) shows the current NAIAD limits
on WIMP-nucleon spin-independent (WIMP-proton spin-dependent)
cross-section as functions
of WIMP mass based on the data described in Table \ref{stat}.
Also shown in Figure \ref{limits}a are the 
region of parameter space favored by the DAMA positive
annual modulation signal (DAMA/NaI-1 through DAMA/NaI-4) 
\cite{dama} (closed curve), limits on the cross-section set by the DAMA
experiment (DAMA/NaI-0) using pulse shape analysis (dashed curve) and
the projected limit of DAMA experiment (DAMA/NaI-0 through
DAMA/NaI-4, dotted curve), if the DAMA group applies pulse shape
discrimination to all available data sets (see \cite{csitest}
for discussion). It is obvious that the DAMA group
could confirm or refute the signal, observed in their annual modulation
analysis, using pulse shape analysis applied to all five
data sets. Note that we do not show here the world best limits
on spin-independent cross-sections set by CDMS-I \cite{cdms}, 
EDELWEISS \cite{edelweiss} and ZEPLIN-I \cite{zeplin}. A paper
on the ZEPLIN-I experiment, which will discuss this further, 
is in preparation.

\begin{figure}[htb]
\begin{center}
\epsfig{figure=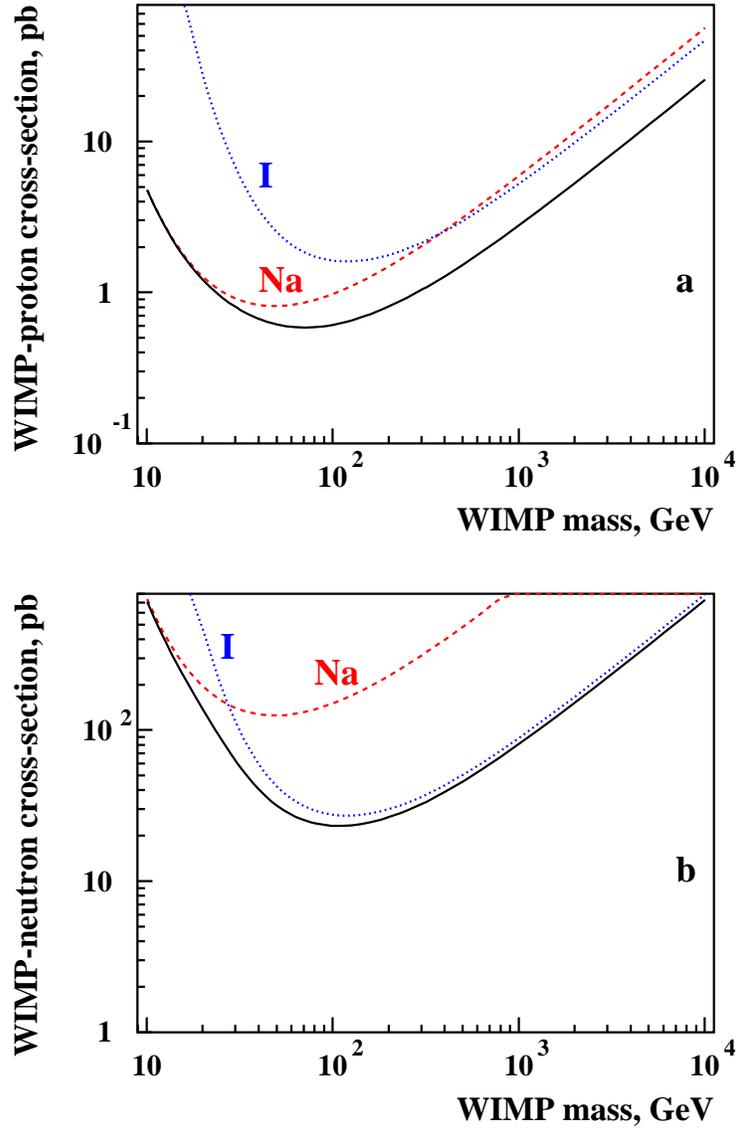,height=16cm}
\caption {NAIAD model-independent limits (90\% C.L.)
on spin-dependent WIMP-proton (a) and WIMP-neutron
(b) cross-sections as functions of WIMP mass.
The limits have been derived following the procedure described
in Ref. \cite{dan3}. Dashed curves show the limits extracted from
interactions with sodium, dotted curves show those from iodine, and
solid curves show combined limits.}
\label{limits1}
\end{center}
\end{figure}

Model-independent limits on spin-dependent WIMP-proton and WIMP-neutron
cross-sections, calculated following the procedure described in 
Ref. \cite{dan3}, are presented in Figure \ref{limits1}. 

A further two crystals have been recently added to the array, increasing 
the total mass to about 65 kg. Recently installed detectors
(unencapsulated DM74 in dry air, encapsulated DM80 and DM81) 
show higher light collection and better discrimination
than earlier modules. The latter two were installed
at the beginning of 2002 and have a light yield of about 
8 photoelectrons/keV. Once sufficient data are collected, these
will be included in further analysis.

The sensitivity of the NAIAD array is currently restricted by
the presence of PMT noise pulses. These pulses occur mainly when
a discharge in the dynodes
of one PMT is seen by both of them. The noise is
reduced by applying asymmetry cuts as described in Section 3
but is still present in the time constant distributions at low
energies (see Figure \ref{tcd}). We plan to investigate this noise
in detail and improve the software cuts to eliminate it.
Our preliminary estimates show that complete suppression
of this noise may result in a factor of 2 improvement in
sensitivity.

The sensitivity of the data presented here is also
limited by the slow DAQ, based on LeCroy oscilloscopes and Labview 
software. This restricts the rate and hence the statistics of 
the calibration runs
performed with gamma-ray sources. (Note that increasing the time spent
on the calibration runs decreases the time available for real data
taking). Thus, the parameters of the time constant distributions are 
known with a finite accuracy, which is taken into account in the
data analysis. Having moved to the fast Acqiris CompactPCI
digitisers we expect to collect more events with gamma-ray and
neutron sources, improving the accuracy of the best fit parameters.

We are also working on improvements to the pulse fitting
procedure, studying more complicated functions to fit the pulse
shapes, as was suggested in \cite{dan1,dan2}. Finally, we are
looking into more sophisticated ways to set limits on the nuclear 
recoil rate using Poisson statistics, rather than $\chi^{2}$ 
analysis, which does not work properly with small number of events.
All improvements in the analysis, including full suppression
of the noise, may increase the NAIAD sensitivity for a given
statistics by a factor of 3.

We expect that changes to the DAQ, improvements in the data analysis
and light collection, suppression of the PMT noise and increased
statistics will result in the improvement of our current 
limits by a factor of 10 in the next 2-3 years of data taking.

\vspace{0.5cm}
{\large \bf 5. Conclusions}
\vspace{0.3cm}

The status of the NAIAD experiment for WIMP
dark matter search at Boulby mine has been presented. 
The detector consists of an array of NaI(Tl) crystals with 
high light yield.
Currently six crystals are collecting data. 
Pulse shape analysis has been used
to discriminate between nuclear recoils, possibly caused
by WIMP interactions, and electron recoils due to
gamma-ray background. We have presented upper limits on the 
WIMP-nucleon spin-independent and WIMP-proton spin-dependent
cross-sections based on the data accumulated by four modules 
(10.6 kg$\times$year exposure).
We expect further improvement in sensitivity by a factor of 10
in the next 2-3 years of data taking based on increased
statistics, increased light yield of new crystals and improved
data analysis.

\vspace{0.5cm}
{\large \bf 6. Acknowledgments}
\vspace{0.3cm}

\indent The Collaboration wishes to thank PPARC for financial support.
We are grateful to the staff of Cleveland Potash Ltd. 
for assistance. We also thank the referee for very useful comments.

\pagebreak

\begin{table}[htb]
\caption{Statistics for NAIAD detectors.}
\vspace{1cm}
\begin{center}
\begin{tabular}{|c|c|c|c|c|}\hline
Crystal & Mass, kg & Light yield, pe/keV & Time, days & Exposure, 
kg$\times$days \\
\hline
DM74 (oil)          & 8.50 & $3.5 \pm 0.4$ & 117.1 & 995.7  \\
DM72                & 3.94 & $7.0 \pm 0.4$ & 274.0 & 1079.4 \\
DM76                & 8.32 & $4.6 \pm 0.3$ & 101.2 &  842.3 \\
DM77                & 8.41 & $6.1 \pm 0.3$ &  62.2 &  522.9 \\
DM74 (dry nitrogen) & 8.40 & $8.4 \pm 0.4$ &  52.2 &  438.7 \\
\hline
Total exposure    &      &               &       & 3879   \\
\hline
\end{tabular}
\end{center}
\label{stat}
\end{table}
\pagebreak

\pagebreak

\pagebreak

\pagebreak

\pagebreak

\pagebreak

\pagebreak

\pagebreak

\pagebreak

\end{document}